# Power Scaling of Narrow-Linewidth Mid-IR Spherical Microlasers

Behsan Behzadi[†], Ravinder K. Jain, Mani Hossein-Zadeh[*]

*Center for High Technology Materials, University of New Mexico, 1313 Goddard SE, Albuquerque, NM 87106*
[†]bbehzadi@unm.edu  [*]mhz@unm.edu

**We discuss issues re: power scaling of mid-IR Er:ZBLAN spherical microlasers and demonstrate achievement of near-milliwatt power levels from a narrow-linewidth ($\Delta\lambda < 50$ pm) 2.71 µm laser source based on an Er:ZBLAN microlaser.**

Narrow linewidth mid-infrared (MIR) lasers are important sources for trace level detection of a broad range of molecular species by spectroscopic methods [1, 2]. For this application, interband and quantum cascade semiconductor lasers (ICLs and QCLs) -- of distributed feedback (DFB) design [3,4] -- are currently the most readily available compact sources of narrow-linewidth mid-IR emission ($\Delta\nu < 30$ MHz). However, the complexity of the growth and fabrication process of these semiconductor lasers makes them relatively expensive, particularly when different designs are needed in small quantities. Moreover, since the circulating optical power is not easily accessible in ICLs and QCLs, these lasers are not easily adaptable for ultrahigh-sensitivity intracavity molecular detection. In addition, the 2.7 - 2.8 µm MIR spectral range for high-sensitivity molecular detection application of several important molecular species (such as $H_2O$, $CO_2$, HF, HOCL) is not easily achievable by narrow linewidth ICLs or QCLs [3,4].

Glass-based whispering gallery mode lasers (WGMLs) are alternative compact sources of narrow linewidth NIR [5,6] and MIR radiation [7,8,9] that are not only highly amenable for intracavity sensing applications (because of the extension of the evanescent tail into the surrounding analyte-bearing medium), but also readily lend themselves to much simpler fabrication processes [5–10], largely because high quality factor ("high-Q") glass WGML microspheres of variable dimensions are relatively easy to fabricate individually, and the doped-glass gain medium also serves as the laser cavity without the need for any external mirror or feedback structures. As such, glass-based mid-IR WGMLs of custom design are anticipated to be less expensive and to generate narrower linewidths (potentially <100 kHz [5]) quite readily.

Despite these clear advantages, WGML sources frequently suffer from the disadvantage of ultralow (µW-level) output powers [5-9], due largely to their small mode areas (A = 5 - 20 µm$^2$), low coupling coefficients (particularly in the case of high-Q operation), and poor pumping efficiencies (due to the inevitably small spatial overlap between the pump and laser modes [9], particularly in MIR WGMLs). These low (µW-level) output powers are a major disadvantage for MIR WGMLs, particularly for sensing applications where detector and amplifier noise can severely reduce the detection sensitivities [1,2,11]. As such, there is a strong need for design and power scaling strategies to boost the powers of WGML-based MIR sources with minimal impact on their linewidths.

In this paper, we discuss power scaling methods and demonstrate near-mW level MIR output powers at 2.71 µm from an amplified Er:ZBLAN MIR WGML. The power scaling was achieved via three steps: 1) improving the output power of the "seed source" mid-IR WGML, 2) designing and fabricating a high-gain Er-Pr:ZBLAN fiber amplifier with a peak gain near the wavelength of the seed laser, and 3) optimizing the power coupling efficiency between the WGML and the amplifier. We show that even after amplification the linewidth of the MIR microlaser stays below $\Delta\lambda \sim 50$ pm ($\Delta\nu \sim 2$ GHz), as needed for high-sensitivity molecular detection applications.

We previously reported the first Er:ZBLAN MIR WGML [8] with an internal slope efficiency, $\eta_{int}$, of 0.35% and an output power ($P_{out}$) of 600 nW at 2 mW input pump power ($P_{in}$) for 320 µW absorbed pump power, $P_{p,Ab}$, using a microsphere (MS) of diam. $D_L$ of 180 µm. Here, $\eta_{int}$ = ratio of the laser output power coupled to the fiber taper at the coupling junction ($P_{L,J}$) and the absorbed pump power

($P_{p,Ab}$), ignoring losses such as attenuation through the fiber coupler and reflection losses that can be remediated quite easily in future designs. A simple theoretical analysis -- based on Ref. 12 -- indicates that the output power, $P_{out}$, for a given pump power, $P_{in}$, was limited primarily by the coupling efficiency between the microsphere and the fiber taper, which resulted in a small external slope efficiency ($\eta_{ext}$) of ~ 0.05%. Here, h$_{ext}$ is the ratio between $P_{L,J}$ and the input pump power at the coupling junction ($P_{P,J}$). For larger microspheres [8], the combination of poor phase matching with WGM modes [9,12] and the small evanescent tail at the pump wavelength results in less efficient pumping. Fig. 1 shows the calculated variation of $\eta_{ext}/\eta_{int}$, which is proportional to the pump absorption efficiency or $P_{p,Ab}/P_{p,J}$ ($P_{p,Ab} = P_{p,T} - P_{p,J}$, where $P_{p,T}$ is the transmitted pump power), as function of the Er:ZBLAN microsphere diam. (with a doping level of 8 mole%). For each data point here, the estimated values are based on the WGM with the largest coupling factor to the fiber taper (of 1 μm diam.) mode; as such, each point may correspond to a different WGM near pump wavelength. As evident from the figure, the pump absorption efficiency for a 180 μm diam. microlaser is about 10 times smaller than its maximum value that occurs around $D_L$ ~ 44 μm, implying preference for using smaller microspheres than the ones we used previously [8,9].

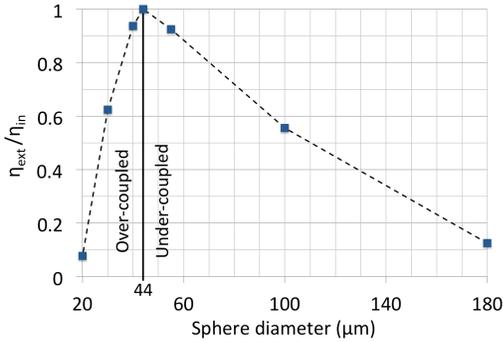

Fig. 1: Plot of the ratio ($\eta_{ext}/\eta_{int}$) of the calculated external and internal slope efficiencies as a function of the Er:ZBLAN microlaser diam. using a silica fiber-taper (of 1 μm diam.) that is in direct contact with the microsphere near its equator. $\eta_{ext}/\eta_{int}=1-(Q_{abs}-Q_{ext}/Q_{abs}+Q_{ext})^2$ where $Q_{abs}$ and $Q_{ext}$ are absorption limited and external quality factors [12] at $\lambda_p$=980 nm respectively (for 8 mole % Er:ZBLAN, $Q_{abs}$ ~ $2\times10^4$).

In our earlier experiments [8,9], fabrication of high-Q ZBLAN microspheres (MSs) was done by using a relatively simple "fiber melting technique" [10] in which the lower limit of the diam. of the resulting MS is determined by the diam. of the available uniformly doped fibers/micro-rods. Since the smallest available Er:ZBLAN fibers/micro-rods had a diam. of ~ 100 μm, the resulting high-Q MSs were typically of D = 150 - 250 μm diam. In order to fabricate microspheres of smaller diam., we reduced the diam., of the original fibers by pulling and tapering these fibers rapidly -- to avoid formation of micro-crystals during the process -- in a custom electric microheater. We were able to fabricate high-Q ZBLAN microspheres with diameters between 55 and 100 μm by melting the tip of such tapered fibers using our original method [10]. Note that while a similar approach has been previously used to fabricate small silica MSs [13], the complex physical characteristics of ZBLAN (especially its tendency for crystallization) has been a major bottleneck for fabrication of small diam. ZBLAN MSs. Fig. 2(a) shows the photograph of an undoped ZBLAN MS with a diam. of ~ 55 μm. Due to the lack of a narrow-linewidth tunable MIR laser for characterization of the Q (quality factor) of these MSs, we inferred the Q's of the small ZBLAN microspheres from the value measured at a wavelength of 1.55 μm. Fig. 2(b) shows the normalized transmission spectrum of such a MS. The "loaded" value of the optical Q of this MS near 1550 nm is $9.2\times10^6$ corresponding to an intrinsic Q > $10^7$ near 2.7 μm (based on the measured loss of ZBLAN fibers made of the same material) [10]. We fabricated several Er:ZBLAN MSs (with uniform Er density of 8 mol %) of diam. between 55 and 150 μm, and characterized their lasing properties using a fiber taper coupler fabricated using a multi-mode low-OH silica fiber and a standard experimental configuration [8,9]. As expected, the largest $\eta_{ext}$ was observed with a microlaser of 55 μm diam. (the smallest MS fabricated).

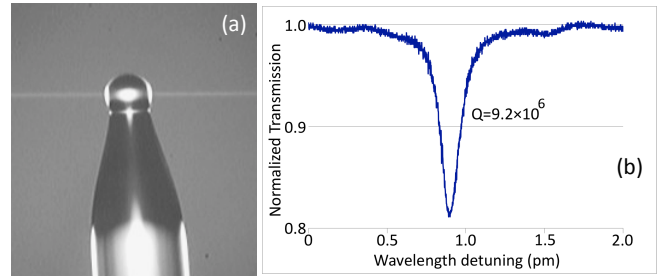

Fig. 2: (a) Photograph of a ZBLAN microsphere with a diameter of 55 μm. (b) Normalized transmission of a taper-coupled tunable 1550 nm in this passive ZBLAN microsphere (for estimating its intrinsic Q).

Fig. 3 shows the $P_{L,J}$ plotted against $P_{P,J}$ for the original MIR WGML (D = 180 μm) reported in Ref. 8 and a WGML of diam. D = 55 μm (this work). In both microlasers, the fiber-taper was in direct contact with the small microlaser ($g = 0$). Note that $P_{L,J}$ was inferred from the transmitted laser power ($P_{L,T}$), taking into account the bidirectional nature of the microlaser and the propagation loss from the junction to the output port. The measured MIR power is still two times smaller than $P_{L,T}$ due to reflection from the dielectric-air interfaces (lens and filter) before the photodetector. Based on our measurements, for the small (D = 55 µm) microlaser (ML), 80% of the input pump power available at the coupling junction was absorbed, and subsequently converted to MIR radiation ($\lambda$=2.71 µm) with an external slope efficiency of 0.29%. As a result of the improved coupling efficiency and better spatial overlap of the pump and laser modes, $\eta_{ext}$ was enhanced by about 6 times in the smaller MS.

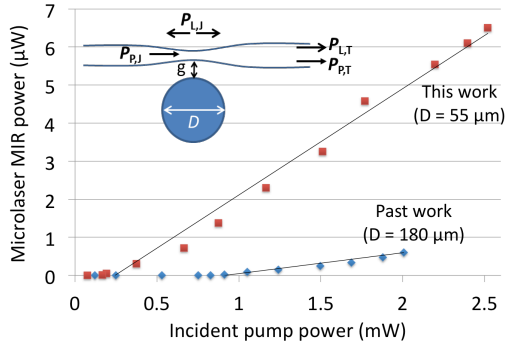

Fig. 3: The total mid-IR output power plotted against incident pump power for a Er:ZBLAN WGMLs: Red squares: diameter=55 µm. Blue squares: diameter=180 µm [8]. The inset shows the definition of the junction and transmitted powers.

Note that the measured value of $\eta_{ext}/\eta_{in}$ (80%) is lower than its theoretical value (~90%, see Fig. 1) because in our calculation we only consider WGMs with maximum coupling to the fiber taper, while in reality the pump may couple to other modes. Moreover the exact diam. of the fiber taper is difficult to measure. Fig. 4(a) shows the variation of $P_{L,J}$ as a function of the absorbed pump power ($P_{p,Abs} = P_{P,T} - P_{P,J}$) for the same ML (see Fig. 2(a)) for zero and non-zero coupling gaps. The internal slope efficiencies ($\eta_{int}$'s) are 0.36% and 0.17% for $g = 0$ and $g > 0$ respectively. As such, for this ML, operation in the "contact mode" not only makes the system more stable -- by eliminating the power fluctuations induced by coupling fluctuations -- but also results in larger optical output power. To show that $g = 0$ correspond to a near-critical coupling condition for the pump wavelength, we measured $P_{L,J}$ and residual pump power ($P_{P,T}$) as a function of g (Fig. 4(b)). Here zero displacement corresponds to the largest gap at which the mid-IR output power is detectable by the photodetector. Moving the microsphere 80 nm closer to the fiber-taper makes the gap small enough for the electrostatic force to pull the fiber-taper toward the microsphere and attach them together (as such smaller coupling gaps could not be measured). The behavior in Fig, 4(b) shows that even for the small MS, the residual pump power decreases (coupled pump power increases) by decreasing the gap ($g$), indicating that the system is getting close to the critical coupling condition, but is still under-coupled at the pump wavelength.

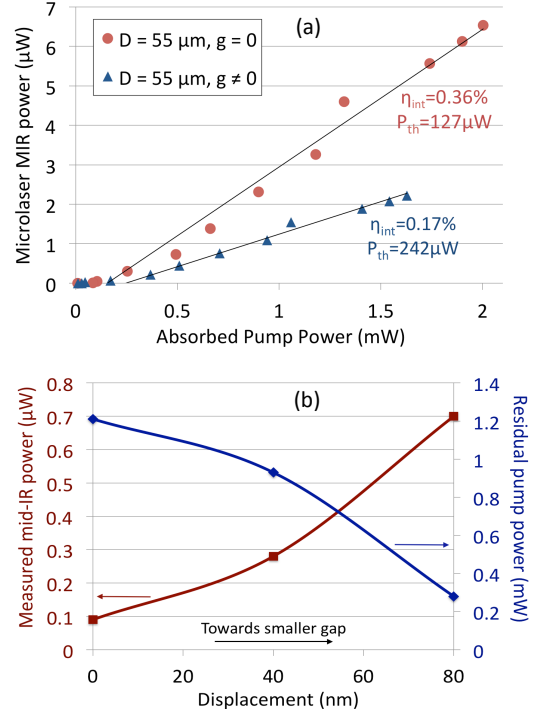

Fig. 4: (a) Plot of the total mid-IR output power for a D = 55 µm) Er:ZBLAN microlaser for two values of gap ($g$) (b) Plot of the mid-IR output power and the residual pump power as a function of the fiber-taper displacement. Note that in Fig. 2 (b), 80 nm displacement corresponds to zero gap (g=0).

The next stage of power scaling requires increasing the output powers to desirable levels with minimal degradation of the key specifications needed. For our application, namely spectroscopic sensing, we focused on demonstrating a mW power level source -- by using an optimized MIR fiber amplifier [14], with a target gain of 30 dB, limited only by the onset of spurious parasitic oscillations [15,16] -- without serious degradation of the amplitude noise or the linewidth of the ML-based mid-IR source. Due to the red-shift [14] of the mid-IR gain at high pump intensities in heavily-doped Er:ZBLAN glasses, the microlaser output spectrum is not well-matched to the gain spectrum of an Er:ZBLAN fiber amplifier whose doping closely matches that of the microsphere. In order to achieve high gain at the ML ouput wavelength (~2.71 µm), we constructed a MIR amplifier, depicted schematically in Fig. 5 below, using a double-clad

"DCF" (13 μm core and 125 μm concentric circular clad) 4.8 m long Er-Pr:ZBLAN [14] fiber (labeled "EPDF" in Fig. 5) with 2 mol% $Er^{3+}$ and 0.5 mol% $Pr^{3+}$ and a 980 nm high power (>5 Watt) laser diode ("980 nm HP DL" in Fig. 5) pump. A low-OH silica fiber, initially of 50 μm core diam. was used for fabrication of the fiber taper (of estimated 1 μm diam. in the MS coupling region). As part of the optimization process for the ML-EPDFA composite system, to ensure efficient coupling between the ML and the EPDFA, we replaced the low-OH fiber-taper with a single-mode (SM) Ge-doped fiber taper (the Ge-doped has a relatively low MIR loss, ~15 dB/m). Note that fabrication of "preferred" fiber tapers couplers -- with glasses of high mid-IR transparency (typically soft glasses) and refractive indices close to that of the ZBLAN MS is quite difficult, and has not been demonstrated by us yet.

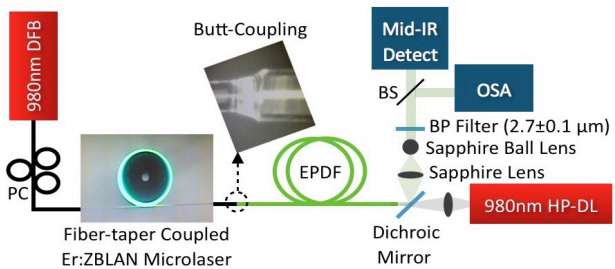

Fig. 5. Experimental arrangement for amplifying Er:ZBLAN microlaser

Although the transmission loss of the Ge-doped fiber used in this setup is ~5 times larger than that of low-OH fiber, its smaller mode field diameter (MFD = 13 μm versus 66 μm) enables more efficient coupling to the core of the EPDFA, whose MFD is ~16 μm. As shown in the photographic "blow-up" of the coupling junction between the Ge-doped output coupler fiber-to-EPDFA input in Fig. 5, this junction was butt-coupled and reflections were minimized by means of an index matching liquid to enable high gain operation and minimize the onset of parasitic oscillations [16]. The distal end of the EPDF was cleaved at an angle of 11° to further inhibit the onset of such spurious oscillations. A dichroic 45° beamsplitter (HR > 99.5% in the MIR and HT > 95% at 980 nm) was used to redirect the MIR output power into a sapphire ball lens (5 mm diameter) to generate a collimated beam into the MIR diagnostic equipment (OSA, detector, power meter) after transiting through a bandpass filter (of BW Δλ ~ 200 nm, $\lambda_o$ ~ 2700 nm) that blocks the residual 980 nm pump radiation.

Fig. 6 shows the measured output power from an EPDFA-amplified Er:ZBLAN ML as a function of the EPDFA pump power. For this figure, the output power of the Er:ZBLAN ML into EPDFA was 1.3 μW, yielding an amplified power of 986 mW (prior to the onset of spurious lasing at a pump power of 4.8 W, see inset), which corresponds to the highest reported gain (28.7 dB gain) for a MIR fiber amplifier.

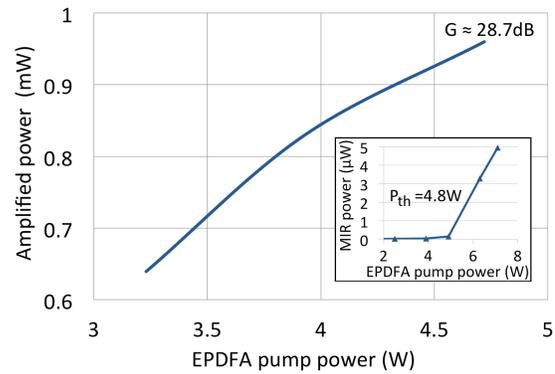

Fig. 6: Output power of the amplified MIR microlaser plotted against amplifier pump power when the coupled microlaser power into the EPDFA is 1.3 μW. Inset: Threshold measurement of EPDFA.

Note that the power delivered to the amplifier is less than the total laser power generated at the coupling junction partly due to loss through the fiber taper, the Ge-doped silica fiber coupling sections, and the coupling loss at the junction with the EPDF. In future work, replacing the Ge-doped fiber-taper with a low loss MIR fiber-taper, with a MIR-transparent fiber spliced to a mode-matched optimally-pumped EPDFA should result in at least a 2-fold increase in the output power of such a MIR source.

Fig. 7 shows the spectrum of the amplified mid-IR source with an OSA resolution-limited measured linewidth of < 50 pm. The measured resolution-limited linewidth of the pre-amplified near-threshold output of the microlaser was also < 50pm (inset of Fig. 7), indicating that the resolution-limited linewidth of the amplified microlaser was not significantly degraded by the amplification process in the EPDFA.

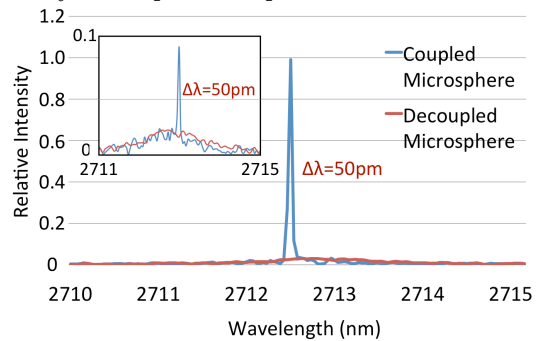

Fig. 7. Spectrum of the amplified mid-IR power (blue line) and corresponding ASE (red line) when the launched power from the microlaser to EPDFA is 142 nW. The inset shows the same spectrums when the microlaser output power is 6 nW.

The fact that the near-threshold (microlaser $P_{out}$ = 6 nW) the linewidth is < 50 pm implies that at larger laser powers the linewidth is significantly smaller. Using the same methodology described in [9], the total quality factor ($Q_{tot}$) of microlaser estimated from the Schawlow-Townes linewidth relation is at least 12,500, and its linewidth expected to be sub-MHz at an output power, $P_{out}$, of 6.5 µW.

In summary, we have demonstrated a narrow linewidth (<50 pm) 2.71 µm laser source with near 1 mW output by combining an optimized Er:ZBLAN WGML (as the seed laser) with an Er-Pr:ZBLAN fiber amplifier. To achieve such a narrow linewidth using conventional MIR fiber lasers is very difficult and requires complex arrangements, careful alignment and employment of expensive high quality mirrors (or gratings). As described above the maximum output power of this source is limited only by the onset of laser oscillation in the amplifier, as is a well-known problem for single-stage fiber amplifiers and can be addressed by employing a multistage MIR fiber amplifier where each stage is isolated from the previous one by an optical isolator. We anticipate the achievement an output power of above 1 W with such a narrow-linewidth MIR source, limited only by the onset of SBS in the amplifier fiber. The performance of the seed microlaser was improved in part by reducing the microsphere diameter such that in the "contact" mode (g=0), its coupling to a robust fiber-taper ( of ~1 µm diam.) is near-critically-coupled for the pump wavelength (980 nm). A simple analysis shows that due to the large difference between the magnitude of the pump and laser wavelengths (a factor of ~ 3), when the pump wavelength is critically coupled, the laser wavelength is over-coupled. In other words, the output power of the microlaser is improved at the expense of the total quality factor ($Q_{tot}$) of the laser. As such, if a narrower linewidth with lower power is preferred in certain applications, one may prefer to use larger microspheres (because the laser linewidth is inversely proportional to $Q_{tot}^2$). In such a case, the ML output power will be even smaller, and it will be even more desirable to amplify the microlaser output power using a fiber amplifier system similar to one demonstrated here.

**Funding.** National Science Foundation (NSF) Grant #1232263.

**Acknowledgment**. We thank Dr. X. Zhu of the Optical Sciences Center at the University of Arizona for loan of the Er-Pr: ZBLAN fiber for preliminary experiments on the EPDFA described here.